\documentclass{article}
\usepackage[T1]{fontenc}
\usepackage[latin9]{inputenc}
\usepackage[letterpaper]{geometry}
\usepackage{textcomp}
\usepackage{amstext}
\usepackage{graphicx}
\usepackage{esint}
\usepackage[authoryear]{natbib}

\makeatletter
\newcommand{\lyxaddress}[1]{
	\par {\raggedright #1
	\vspace{1.4em}
	\noindent\par}
}

\makeatother

\begin{document}

\title{Determining the depth of Jupiter's Great Red Spot with Juno: a Slepian approach}
\author{Eli Galanti$^{1}$, Yohai Kaspi$^{1}$, Frederik J. Simons$^{2}$, Daniele Durante$^{3}$, Marzia Parisi$^{4}$,\\  and Scott J. Bolton$^{5}$ \\
\\
(Astrophysical Journal Letters, in press)}
\maketitle

\lyxaddress{\begin{center}
\textit{$^{1}$Department of Earth and Planetary Sciences, Weizmann
Institute of Science, Rehovot, Israel }\\
\textit{$^{2}$Department of Geosciences, Princeton University, New Jersey, USA}\\
\textit{$^{3}$Dipartimento di Ingegneria Meccanica e Aerospaziale,
Sapienza Università di Roma, Rome, Italy}\\
\textit{$^{4}$Jet Propulsion Laboratory, California Institute of Technology, Pasadena, California , USA}\\
\textit{$^{5}$outhwest Research Institute, San Antonio, Texas , USA}
\par\end{center}}

\begin{abstract}
One of Jupiter's most prominent atmospheric features, the Great Red
Spot (GRS), has been observed for more than two centuries, yet little
is known about its structure and dynamics below its observed cloud-level.
While its anticyclonic vortex appearance suggests it might be a shallow
weather-layer feature, the very long time span for which it was observed
implies it is likely deeply rooted, otherwise it would have been sheared
apart by Jupiter's turbulent atmosphere. Determining the GRS depth
will shed light not only on the processes governing the GRS, but on
the dynamics of Jupiter's atmosphere as a whole. The Juno mission
single flyby over the GRS (PJ7) discovered using microwave radiometer
measurements that the GRS is at least a couple hundred kilometers
deep \citep{Li2017a}. The next flybys over the GRS (PJ18 and PJ21),
will allow high-precision gravity measurements that can be used to
estimate how deep the GRS winds penetrate below the cloud-level. Here
we propose a novel method to determine the depth of the GRS based
on the new gravity measurements and a Slepian function approach that
enables an effective representation of the wind-induced spatially-confined
gravity signal, and an efficient determination of the GRS depth given
the limited measurements. We show that with this method the gravity
signal of the GRS should be detectable for wind depths deeper than
300 kilometers, with reasonable uncertainties that depend on depth
(e.g., $\pm$100~km for a GRS depth of 1000~km).
\end{abstract}

\section{Introduction}

Jupiter's Great Red Spot (GRS) has been an iconic feature in the Solar
System for centuries. Ever since it was discovered, hundreds of years
ago, it perplexed astronomers with its shape, color and consistency.
Nonetheless, little is known about the GRS, particularly about how
deep into the gaseous planet this anticyclonic vortex extends. On
the one hand, it resembles an Earth-like atmospheric vortex, suggesting
it is driven by shallow atmospheric processes, and should be shallow
and confined to some weather-layer \citep{Dowling1988,Dowling1989}.
On the other, its centuries-long existence within Jupiter'
's turbulent
atmosphere indicates that it must contain significant mass otherwise
it would have been sheared apart by the jets and other vortices. The
depth to which it extends carries with it great implications on the
mechanisms driving and maintaining it. Until recently, the depth of
Jupiter's atmosphere itself was unknown, but recent gravity measurements
by the Juno spacecraft \citep{Iess2018a} allowed determining that
the atmospheric jets on Jupiter extend down to depths of thousands
of kilometers ($\sim10^{5}$ bars in pressure, \citealt{Kaspi2018}).
The goal of this study is to propose a new methodology for interpreting
the Juno gravity measurements in order to determine the depth of the
GRS.

\begin{figure*}[t]
\centering{}\includegraphics[scale=0.4]{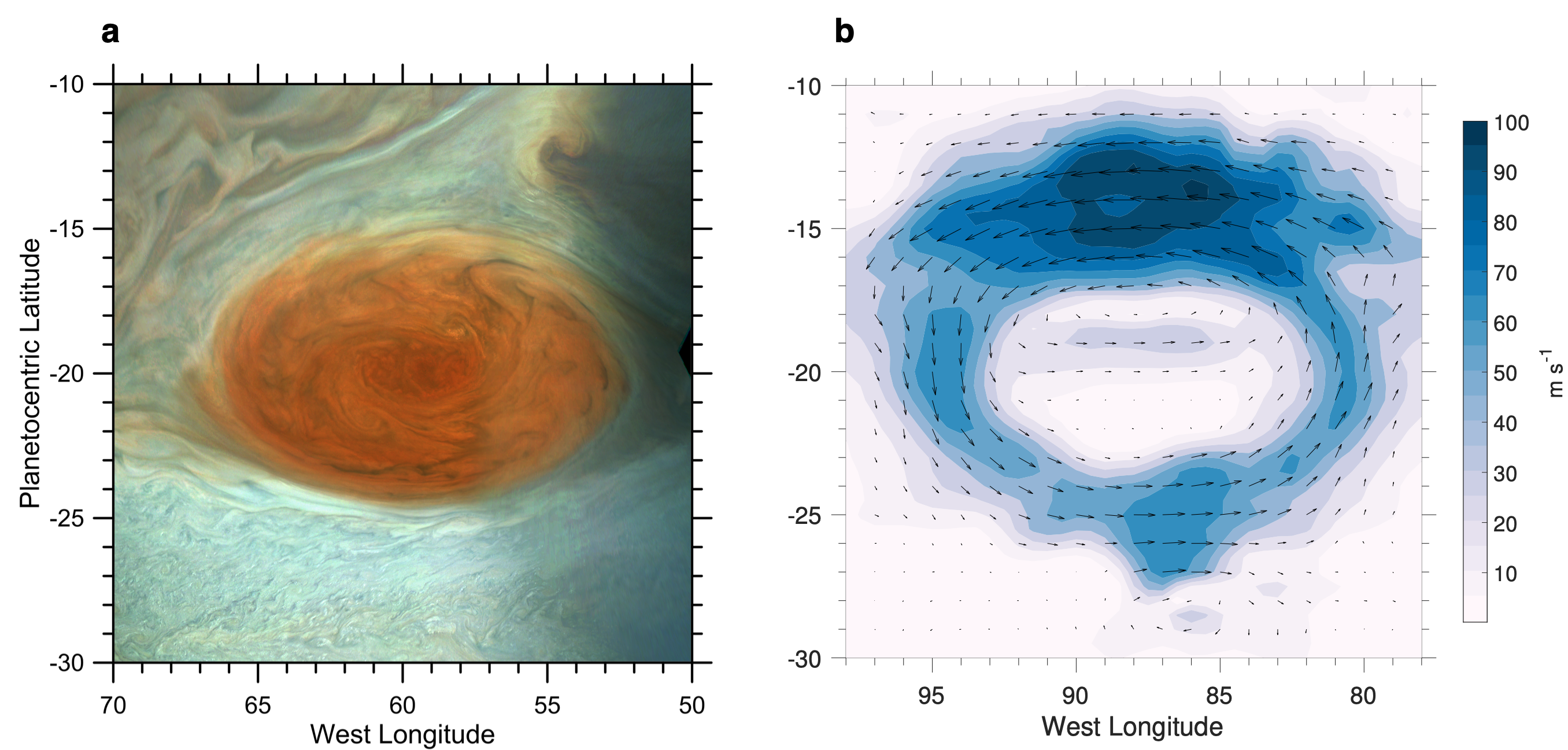}\caption{(a) A JunoCam picture of the GRS obtained on July 17, 2017\citep[Figure~4 from ][]{Sanchez-Lavega2018}.
(b) The non-zonally averaged winds (wind vectors in arrows and magnitude
in colors) in the GRS region based on the 2000 Cassini flyby \citep{Choi2011}.\label{fig: GRS location}}
\end{figure*}

The Juno spacecraft orbits Jupiter every 53 days on a polar, highly
eccentric orbit with perijove at 4000~km above Jupiter's cloud-level
\citep{Bolton2017,Folkner2017}. To allow a full coverage of the planet,
every perijove is at a different longitude with a planned longitudinal
separation of 11.25 degrees over the entire mission. As the GRS drifts
by $\sim$110 degrees eastward every year \citep{Simon2018} this
provides, in principle, several opportunities for passes over the
GRS. The first of these has been on orbit 7 (PJ7), but this orbit
was devoted to microwave measurements, meaning radio science operated
only in the X-band, thus preventing applying the plasma calibration
scheme, which requires simultaneous Ka-band data \citep{Iess2018a}.
Orbits 18 (PJ18) and 21 (PJ21) will fly over the GRS in gravity mode,
meaning they will operate with the more accurate Ka-band.

Differently than the depth estimate of the zonal jets, which is obtained
using the zonal gravity harmonics, for the case of the GRS a non-zonal
localized field is required. \citet{Parisi2016} used the tesseral
gravity field to estimate the depth, and found that the GRS must be
at least 2000~km deep in order to be detected. However, that estimate
of the gravity signature required the determination of a large number
of spherical harmonics, resulting in a considerable uncertainty in
the solution. Here we propose a new approach, using Slepian functions
that are designed specifically for isolated gravity measurements of
local spatial features \citep[e.g.,][]{Simons2006b,Simons2009,Harig2012,Plattner2017}.
We demonstrate its applicability to the GRS problem and examine the
detectability of the GRS depth with the method, given the limited
measurements expected.

\section{Methods}

\subsection{Definition of Slepian functions}

Given a phenomenon, such as the GRS winds and its accompanying gravity
field, that is confined to a specific region, the Slepian functions
form a basis set specifically to maximize the phenomenon representation
inside the region, and minimize it outside the region. This is in
contrast to the traditional spherical harmonics that are set to represent
global signals with no preferences to specific regions. We follow
here the derivation given by \citet{Simons2006}, sections 3.3 and
4.1, for Slepian functions concentrated over an arbitrarily shaped
region constructed from a band-limited set of spherical harmonic functions.
Let $g(\hat{r})$, a real-valued function on a unit sphere, be given
by a spherical harmonic expansion to bandwidth $L$,
\begin{equation}
g=\sum_{l=0}^{L}\sum_{m=-l}^{l}g_{lm}Y_{lm},\,\,\,\,\,\,\,g_{lm}=\int_{\Omega}gY_{lm}\,d\Omega,\label{eq:function definition}
\end{equation}
where $l$ and $m$ are the degree and order, and $\Omega$ is the
area of the sphere. Defining the spatial and spectral norms as
\begin{equation}
\Vert g\Vert{}_{R}^{2}=\int_{R}g^{2}d\Omega,\,\,\,\,\,\,\,\,\,\,\Vert g\Vert{}_{L}^{2}=\sum_{l=0}^{L}\sum_{m=-l}^{l}g_{lm}^2,\label{eq:norms}
\end{equation}
where $R$ is the region of interest, the problem of maximizing the
spatial concentration of $g$ becomes
\begin{equation}
\lambda=\frac{\Vert g\Vert{}_{R}^{2}}{\Vert g\Vert{}_{\Omega}^{2}}=\frac{\int_{R}g^{2}\,d\Omega}{\int_{\Omega}g^{2}\,d\Omega}={\rm maximum.}
\end{equation}
The ratio $0<\lambda<1$ is a measure of the spatial concentration.
Using Equations~(\ref{eq:function definition}) and (\ref{eq:norms})
this measure can be written as
\[
\lambda=\frac{\sum_{l=0}^{L}\sum_{m=-l}^{l}g_{lm}\sum_{l'=0}^{L}\sum_{m'=-l'}^{l'}D_{lm,l'm'}g_{l'm'}}{\sum_{l=0}^{L}\sum_{m=-l}^{l}g_{lm}^{2}},
\]
where
\[
D_{lm,l'm'}=\int_{R}Y_{lm}Y_{l'm'}\,d\Omega.
\]
The problem can now be formulated as an $(L+1)^{2}\times(L+1)^{2}$
algebraic eigenvalue problem 
\begin{equation}
{\bf Dg}=\lambda{\bf g}.\label{eq:eigenvalue problem}
\end{equation}
The solutions to this equation form the Slepian basis of concentrated
functions in the region $R$. Each solution $\mathbf{g}$ is a vector
including the amplitudes of each of the $(L+1)^{2}$ spherical harmonics
used in the definition. The number of meaningful functions (that are
well concentrated in the region $R$ of area $A$) depends on the
bandwidth and the fractional area of the region of concentration,
and can be calculated using the Shannon number 
\begin{equation}
N=\sum_{i=1}^{(L+1)^{2}}\lambda_{i}=(L+1)^{2}\frac{A}{4\pi},\label{eq:Shannon number}
\end{equation}
where $\lambda_{i}$ are the solutions to Equation~(\ref{eq:eigenvalue problem}).

We can now define the basis of Slepian functions for the region of
interest. We define it as an ellipse centered at $273$$^{\circ}$E
and $16$$^{\circ}$S that spans 20 degrees in longitude and 30 degrees
in latitude. This is similar in longitudinal range to the region of
strong winds (Figure~\ref{fig: GRS location}b), but is larger in
latitudinal range and somewhat shifted northward to account for the
actual gravity signal (discussed in section \ref{sec:Realizations-of-the-GRS}).

\begin{figure*}[t]
\centering{}\includegraphics[scale=0.45]{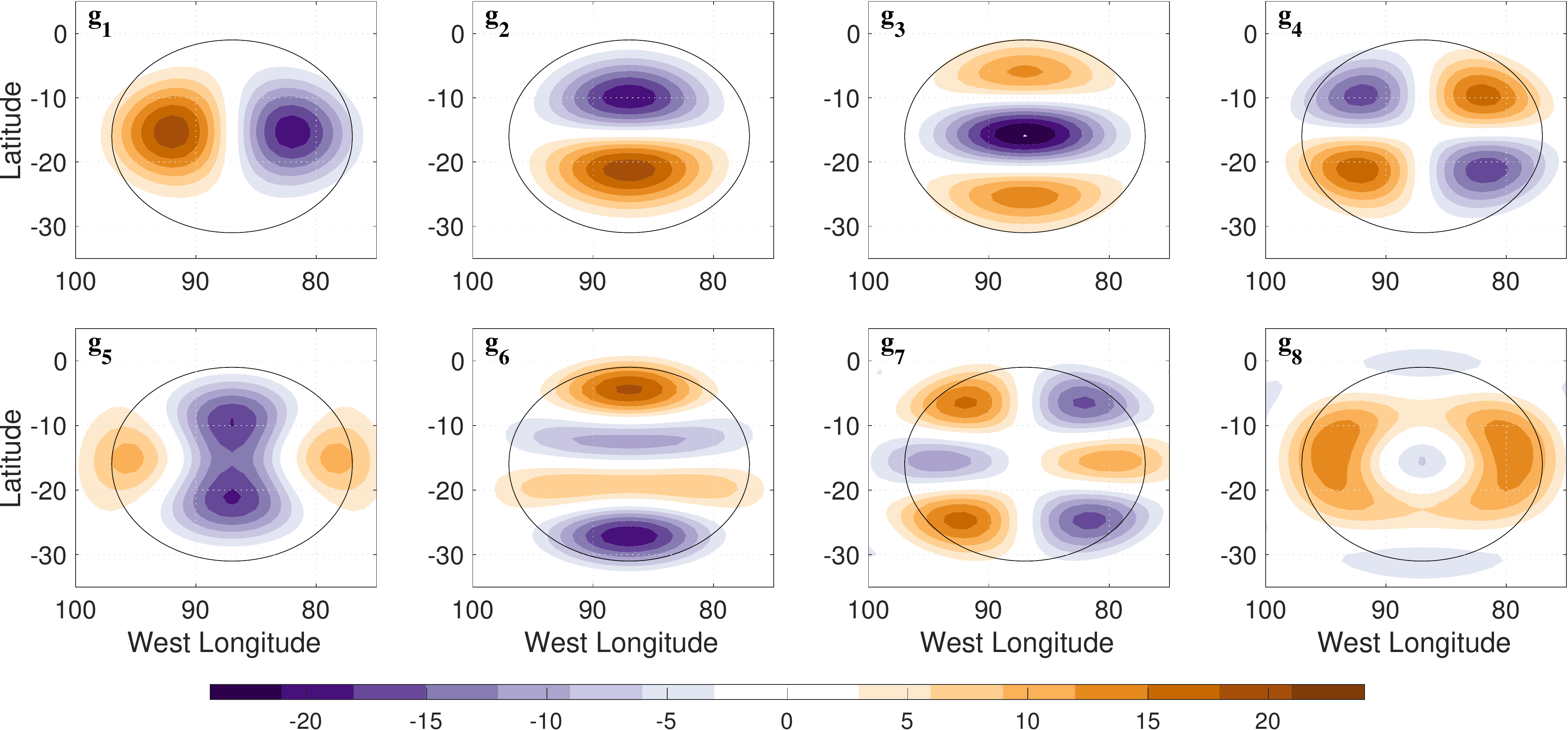}\caption{The first 8 Slepian functions (${\bf g_{1}-g_{8}}$) defined for the
region of the GRS (marked by the black ellipse).\label{fig: Slepian functions-selective}}
\end{figure*}

Equations (1)-(5) define the canonical Slepian functions of \citet{Simons2006},
but here we restrict the range of all of the sums to only include
harmonics with $l=1$ through $l=30$ and $m=1$ through $m=l$. Furthermore,
as the zonal harmonics ($m=0$ for all $l$) are used in the Juno
gravity analysis we exclude them from the Slepian functions. We also
exclude all $m$ for $l=1$ through $l=4$ because the low degree
tesseral field may be related to large scale structure and might be
needed for the overall Juno analysis. This ensures that when incorporating
the GRS Slepian functions in the Juno gravity analysis, there will
be no ambiguity in the values of the spherical harmonics. Our Slepian
functions are selectively band-passed rather than band-limited, and
they remain mutually orthonormal as well as orthogonal to the spherical
harmonics of the excluded degrees and orders, avoiding contamination
with their previously determined coefficients. The calculated first
8 Slepian functions are shown in Figure~\ref{fig: Slepian functions-selective},
and with the Shannon number being $N=10$ we will use in the analysis
the first 10 functions to reconstruct the GRS gravity field.

\subsection{The GRS induced gravity signal\label{subsec:The-TW-method}}

We define the flow structure involved in the GRS similarly to \citet{Parisi2016}
and \citet{Galanti2017c} to be
\begin{eqnarray}
{\bf u}(r,\theta,\phi) & = & {\bf u}_{{\rm cyl}}(r,\theta,\phi)\exp\left[-\frac{a-r}{H}\right],\label{eq:wind-projection}
\end{eqnarray}
where ${\bf u}_{cyl}=[u_{{\rm cyl}}(r,\theta,\phi),v_{{\rm cyl}}(r,\theta,\phi)${]}
is the observed cloud-level wind (Figure~\ref{fig: GRS location}b),
projected along cylinders after its zonal mean is subtracted, $a=69,911$~km
is the planet mean radius, and $H$ is the exponential decay scale
of the cloud-level wind. Note that the exact nature of the decay function
might change for the analysis of the actual measurements \citep[e.g.,][]{Kaspi2018}.

Due to the strong winds around the vortex the geostrophic gradient
creates a density anomaly with respect to its surroundings, that can
be calculated via thermal wind balance, namely
\begin{equation}
\left(2\mathbf{\Omega}\cdot\mathbf{\nabla}\right)\left[\widetilde{\rho}\mathbf{u}\right]=\nabla\rho'\times\mathbf{g_{\mathbf{0}}},\label{eq: thermal wind}
\end{equation}
where $\mathbf{u}(\mathbf{r})=[u_{0},v_{0}]$ is the 3D velocity,
$\mathbf{\Omega}$ is the planetary rotation rate, $\widetilde{\rho}(r)$
is the background density field, $\mathbf{g_{0}}\left(r\right)$ is
the mean gravity vector and $\rho'\left(r,\theta,\phi\right)$ is
the dynamical density anomaly \citep{Pedlosky1987,Kaspi2009}. Other
effects not included in this balance, such as the anomalous gravity
and centrifugal forces induced by the density anomalies \citep{Zhang2015,Cao-Stevenson-2017b},
were shown, for the large scale stronger zonal flows, to have a small
effect on the gravity solutions \citep{Galanti2017a,Kaspi2018}. In
the case of the GRS winds, where the zonally mean wind is excluded,
this holds even more so as the induced density anomalies are local
to the GRS region and the induced gravity anomalies are negligible.
The balance also does not include the effect of the centrifugal force
acting due to curvature of the flow (gradient flow, \citealt{Holton2004}).
This effect implies, for the GRS winds, an increase of about 5\% in
the accompanying density anomalies.

The gravity signal at the surface of the planet resulting from the
density perturbations $\rho'$ can be calculated \citep{Galanti2017c}
either directly or using spherical harmonics coefficients
\begin{eqnarray}
C_{lm}^{{\rm TW}} & = & \frac{1}{Ma^{l}}\frac{2(l-m)!}{(l+m)!}\intop_{r=0}^{a}r^{l+2}dr\nonumber \\
 &  & \times\intop_{\phi=0}^{2\pi}\intop_{\mu=-1}^{1}P_{lm}\left(\mu\right)\cos\left(m\phi\right)\rho'd\mu d\phi,\label{eq:Clm_integration}
\end{eqnarray}
\begin{eqnarray}
S_{lm}^{{\rm TW}} & = & -\frac{1}{Ma^{l}}\frac{2(l-m)!}{(l+m)!}\intop_{r=0}^{a}r^{l+2}dr\nonumber \\
 &  & \times\intop_{\phi=0}^{2\pi}\intop_{\mu=-1}^{1}P_{lm}\left(\mu\right)\sin\left(m\phi\right)\rho'd\mu d\phi,\label{eq:Slm_integration}
\end{eqnarray}
where $l\geq2$ and $m\geq0$ are the degree and order of the expansion,
respectively. The spatially dependent gravity field in the radial
direction is then
\begin{eqnarray}
\delta g_{r}(\mu,\phi) & = & -\frac{GM}{a^{2}}\sum_{l}\left(l+1\right)\nonumber \\
 & \times & \sum_{m=0}^{l}P_{lm}\left(\mu\right)\left[C_{lm}^{{\rm TW}}\cos m\phi+S_{lm}^{{\rm TW}}\sin m\phi\right].\label{eq:gravity_spatial}
\end{eqnarray}

\subsection{Using Slepian functions to define the GRS gravity}

Given a set of Slepian functions $g_{i}$ (Equation~\ref{eq:eigenvalue problem}),
expressed as a combination of spherical harmonics $C_{lm}^{g_{i}}$
and $S_{lm}^{g_{i}}$, and a thermal wind solution for the GRS gravity
signal defined by $C_{lm}^{{\rm TW}}$ and $S_{lm}^{{\rm TW}}$ (Equations~\ref{eq:Clm_integration}
and \ref{eq:Slm_integration}), the combination of the Slepian functions
that best describe the GRS gravity field can be found by minimizing
\[
\mathcal{R}=\sum_{l,m}\left(C_{lm}^{{\rm TW}}-\sum_{i}\alpha_{i}C_{lm}^{g_{i}}\right)^{2}+\sum_{l,m}\left(S_{lm}^{{\rm TW}}-\sum_{i}\alpha_{i}S_{lm}^{g_{i}}\right)^{2},
\]
where $\alpha_{i}$ (the amplitude of the Slepian functions) are the
parameters to be optimized. A solution for $\alpha_{i}$ is found
by solving
\begin{equation}
A\alpha=B,\label{eq:GRS-slepians}
\end{equation}
where
\[
A_{i,j}=\sum_{l,m}\left(C_{lm}^{g_{i}}C_{lm}^{g_{j}}+S_{lm}^{g_{i}}S_{lm}^{g_{j}}\right)
\]
and
\[
B_{j}=\sum_{l,m}\left(C_{lm}^{{\rm TW}}C_{lm}^{g_{j}}+S_{l,m}^{{\rm TW}}S_{lm}^{g_{j}}\right).
\]
Therefore, given a gravity field concentrated in the GRS region (expressed
in terms of spherical harmonics coefficients $C_{lm}^{{\rm TW}},\,\,\,S_{lm}^{{\rm T{\rm W}}}$)
that field can be represented by a set of Slepian functions $C_{lm}^{g_{i}},\,\,\,S_{lm}^{g_{i}}$
weighted by the coefficients $\alpha_{i}$.

The Slepian functions are defined at the planet's surface, but in
practice they have to be projected upward to the location of the Juno
trajectory, where the gravitational pull on the spacecraft is acting.
Theoretically, this could lead to degradation in the orthogonality
of the Slepian functions \citep{Simons2006b,Plattner2017}. The degree
of the degradation is a function of the ratio between the target altitude
and the radius of the planet, as well as the number and complexity
of the Slepian functions used. For the case of the Juno trajectory
over the GRS the altitudes of relevance range are from around 4000~km
at perijove to around 15,000~km. Only at the outer edge of this range
is the uncertainty in the measurement expected to surpass the gravity
signal generated even in cases of very deep winds. The projection
of gravity signal resulting from both the TW solution and that reconstructed
using the Slepian functions can be calculated by multiplying each
coefficient $C_{l,m}^{g_{i}}$, $S_{l,m}^{g_{i}}$ and $C_{l,m}^{{\rm TW}}$,
$S_{l,m}^{{\rm TW}}$ by the factor $\left(\frac{a}{a+h}\right)^{l+2}$,
where $h$ is the altitude to which the gravity field is projected
and $l$ is the degree of the spherical harmonics. Performing this
analysis for the range of wind depths discussed here shows that for
all altitudes the orthogonality of the Slepian functions does not
degrade substantially. In fact, the major error in estimating the
GRS depth with the Slepian functions comes from the number of spherical
harmonics used to define the GRS region, and of course the largest
errors arise from the spatially limited Juno measurements.

\begin{figure*}
\centering{}\includegraphics[scale=0.8]{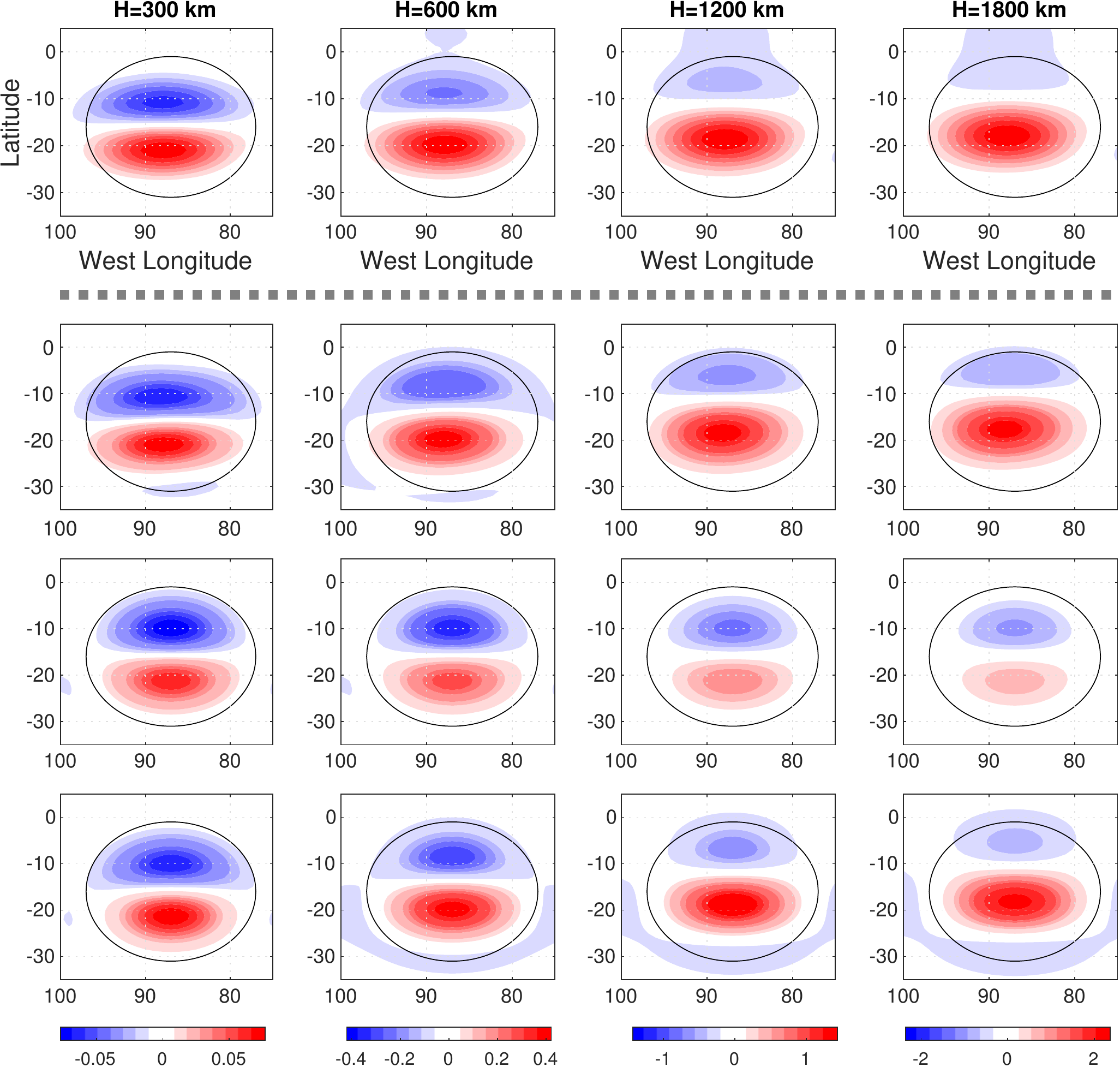}\caption{The gravity signal at the planet's surface (in mGal) resulting from
the GRS winds extended to depths of 300, 600, 1200 and 1800 km. Note
that each case (column-wise) has a different color range. Shown are
the TW solutions (top row), and a reconstruction using all the Slepian
functions $\alpha_{1},\alpha_{2},...,\alpha_{10}$ (second row), $\alpha_{2}$
only (third row), and $\alpha_{2},\alpha_{3,}\alpha_{5},\,{\rm and}\,\alpha_{8}$
(bottom row). The black oval contour denotes the region used to define
the Slepian functions.\label{fig: gravity-TW}}
\end{figure*}

\subsection{Estimating the GRS depth with the trajectory estimation (TE) model}

Juno's radio-science instrumentation is capable of providing very
accurate Doppler measurements, with accuracies as low as 10 micron/s,
at an integration time of 60~seconds. The Doppler measurements are
then analyzed with MONTE, JPL's orbit determination code, to determine
parameters of Juno's dynamical model, such as Jupiter's spherical
harmonics or the Slepian coefficients \citep{Evans2016}.

To assess Juno's sensitivity to the gravitational signal induced by
the GRS, we simulate Juno's gravity experiment up until the end of
the mission. We include all the designed Juno's gravity passes, with
the inclusion of PJ7 (non gravity-dedicated). For the determination
of the GRS gravity signal, the largest contribution comes from the
passes which fly over the GRS, namely PJ7, and the gravity-dedicated
PJ18 and PJ21. We include all the passes planned until the end of
the mission because, in order to be able to determine the signal from
the GRS, a good knowledge of the zonally-symmetric field is required.
The inclusion of the Slepian functions within the orbit determination
code is straightforward since each Slepian function is a known linear
combination of spherical harmonics. Note that the partial derivative
of the observables with respect to the Slepian coefficient $\alpha_{i}$
is also a linear combination of the partial derivatives of the observables
with respect to the spherical harmonics \citep{Han2008,Goossens2012}.

The simulations have been performed by generating 1-day trajectories
around Jupiter's closest approach. Jupiter's gravity field is assumed
to be composed from a zonally symmetric field (following \citealt{Iess2018a})
plus the addition of the signal of the GRS, for the different depths,
represented with the Slepian functions.

\section{Realizations of the GRS induced gravity field\label{sec:Realizations-of-the-GRS}}

The GRS is constantly moving in the zonal direction with respect to
Jupiter system III with a rate of about $0.3\,^{\circ}/{\rm day}$
\citep{Simon2018}, and therefore its longitudinal position has to
be determined with respect to PJ18 and PJ21. Other changes to the
GRS exists with time, such as its size and strength, but these should
not change much within the time interval between the time in which
the GRS shape is determined and the time of the gravity measurements.
An example of the GRS, as observed recently in the cloud-level \citep[Fig. 4 from][]{Sanchez-Lavega2018},
is shown in Figure~\ref{fig: GRS location}a. The strongest winds
are expected in the transition between the brown (belts) and white
(zones) clouds.

\begin{figure}[th]
\centering{}\includegraphics[scale=0.58]{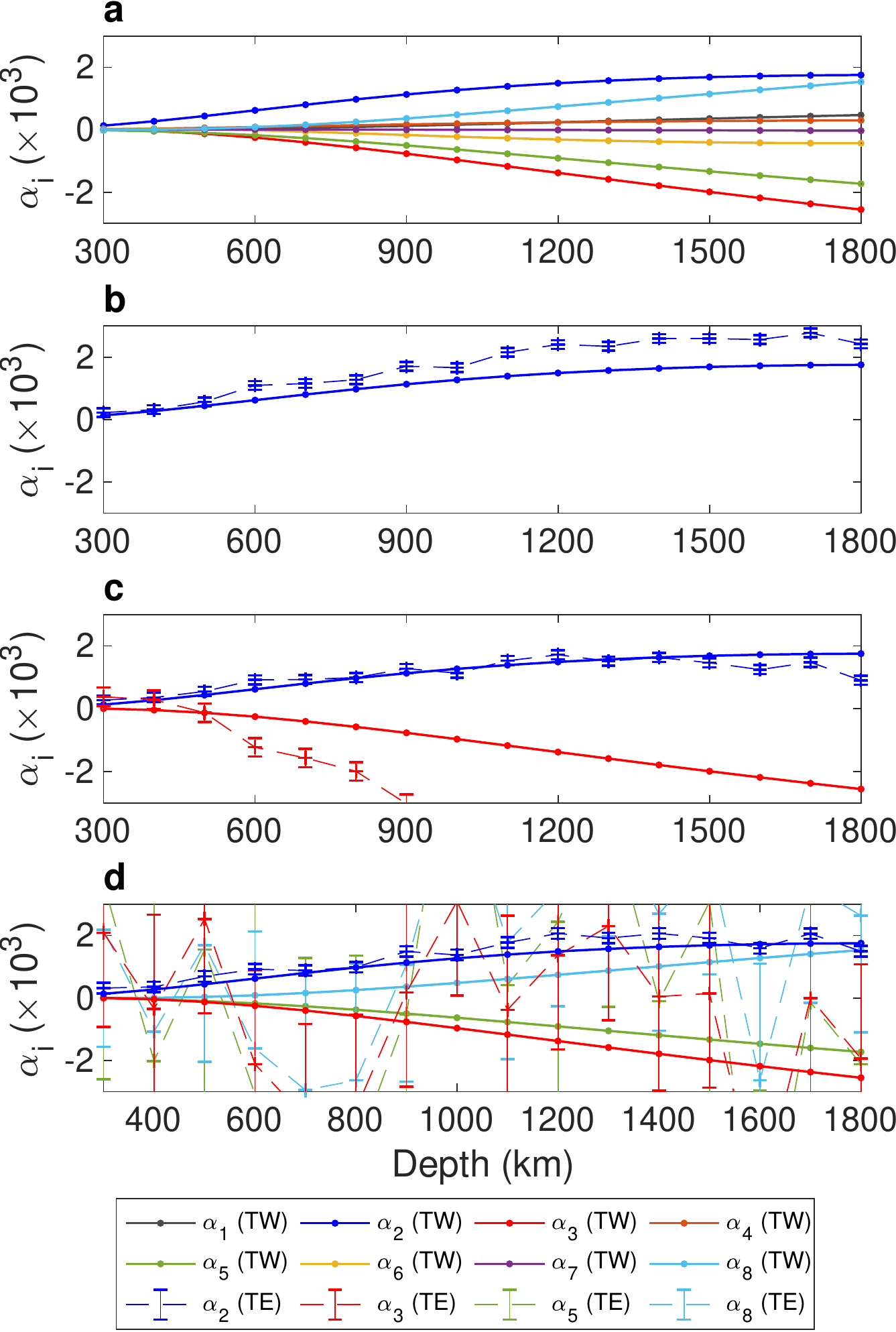}\caption{(a) The $\alpha$ values (solution of Eq.~\ref{eq:GRS-slepians})
$\times10^{3}$) as function of the GRS depth, for the first 8 Slepian
functions used to reconstruct the gravity signals shown in Fig.~\ref{fig: gravity-TW}.
(b) The same value of $\alpha_{2}$ (solid) together with the TE solutions
for it when only $\alpha_{2}$ is optimized for (dashed). Also shown
are the $1\sigma$ uncertainties of the TE solutions. (c) Similar
to (b), but for the TE analysis when both $\alpha_{2}$ and $\alpha_{3}$
are optimized for. (d) Similar to (b), but for the TE analysis when
$\alpha_{2},\alpha_{3,}\alpha_{5},\,{\rm and}\,\alpha_{8}$ are optimized
for.\label{fig: alpha values}}
\end{figure}

The GRS-induced gravity signal is calculated as defined in Section~\ref{subsec:The-TW-method},
thus allowing the examination of the expected gravity signal under
different scenarios of wind depths. The calculated gravity is shown
in Figure~\ref{fig: gravity-TW} (top panels) for e-folding depths
of 300, 600, 1200 and 1800 km. For shallow winds, the signal is mostly
a dipole in the north-south direction with a negative patch north
of the positive one. The dipole structure results from a combination
of factors, coming from the product of the background density and
the winds, on the $lhs$ of Equation~\ref{eq: thermal wind}, since
the former increases with depth and the latter decreases with depth.
As a result, the gradient in the direction of the axis of rotation
gives positive values in the upper layers and negative values in the
deeper layers, thus creating a dipole structure in the radial direction.
However, because the winds extend inward in the direction of the axis
of rotation (Equation~\ref{eq: thermal wind}), this vertical dipole
is shifted such that the negative density anomalies in the deeper
layers are seated northward of the positive density anomalies in the
upper layers. This slantwise density dipole, when integrated (Equations~\ref{eq:Clm_integration},\ref{eq:Slm_integration}),
results in a north-south dipole in the gravity anomalies. For deeper
winds, the positive part of the gravity dipole becomes stronger compared
to the negative part, with a slight shift of the entire pattern to
equatorward The weakening of the negative part of the dipole results
from the negative density anomalies being pushed into deeper layers
that have less effect on the surface gravity anomalies.

With the Slepian functions we can reconstruct the gravity field for
the 4 wind depth cases (Figure~\ref{fig: gravity-TW}, upper panels)
using Equation~(\ref{eq:GRS-slepians}). First, we reconstruct it
with all 10 Slepian functions (Figure~\ref{fig: gravity-TW}, second
row panels). It is evident that most of the gravity signal is being
reconstructed with these functions. The $\alpha_{i}$ values calculated
for the first 8 Slepian functions are shown in Figure~\ref{fig: alpha values}a,
for depth ranging from 300 to 1800~km. Several characteristics appear:
first, there are 4 functions that determine most of the signal: $\alpha_{2},\alpha_{3,}\alpha_{5},\,{\rm and}\,\alpha_{8}$.
Second, for shallow flows the largest contribution comes from $\alpha_{2}$
(blue), but for cases with winds deeper than 700\ km other functions
make a sizable contribution as well, mostly $\alpha_{3}$ (red), $\alpha_{5}$
(green), and $\alpha_{8}$ (light blue). Given this behavior, we can
reconstruct the gravity signal with a subset of the functions. In
Figure~\ref{fig: gravity-TW} (third and forth rows) we show 2 cases
- using $\alpha_{2}$ only, and using $\alpha_{2},\alpha_{3,}\alpha_{5},\,{\rm and}\,\alpha_{8}$.
As expected, using only $\alpha_{2}$ results in a fairly good reconstruction
for the shallower cases, while using the 4 leading Slepian functions
results in a reconstruction that is very similar to the original signals.

\section{The GRS detectability}

We perform a theoretical examination of the method, using the planned
trajectories of PJ18 and PJ21, similar to the analysis of \citet{Galanti2017c},
where the wind-induced gravity field is used to simulate the Juno
trajectories with the Trajectory Estimation (TE) model. Here, we use
the wind-induced gravity signal in the GRS region (Figure~\ref{fig: gravity-TW})
for a depth range of 300-1800~km to simulate the expected effect
on the Juno trajectories PJ18 and PJ21. Then, given the modified Juno
trajectory, we include the Slepian functions in the gravity analysis,
to examine if the wind-induced values for the Slepian functions $\alpha_{i}$
(Figure~\ref{fig: alpha values}a) can be recovered.

Reconstructing with all 10 Slepian functions turns out to be not feasible
since the uncertainties associated with the solution are much larger
than the values themselves, due to large correlations, forcing a reduction
in the number of functions to be used. Identifying the most important
functions for the GRS gravity signal (largest values in Figure~\ref{fig: alpha values}a)
we can use only a subset of the functions in the estimation process.
The largest contribution comes from $\alpha_{2}$, especially for
the shallow cases, since using the TE model to fit the gravity field
with $\alpha_{2}$ only (Figure~\ref{fig: alpha values}b, dashed)
shows a fairly good fit for all cases with some overestimation for
deep flows. Adding $\alpha_{3}$ as a second parameter to the fit
(Figure~\ref{fig: alpha values}c, dashed) shifts $\alpha_{2}$ to
give very good values even for deep winds, but the value of $\alpha_{3}$
cannot be recovered for winds deeper than 500\ km. Finally, fitting
with $\alpha_{2}$, $\alpha_{3}$, $\alpha_{5}$, and $\alpha_{8}$
(Figure~\ref{fig: alpha values}d, dashed) gives again a good estimate
for $\alpha_{2}$, but the other 3 Slepian functions remain unresolved,
suggesting that aside from $\alpha_{2}$ the other Slepian functions
are highly correlated in their manifestation in the Juno trajectory.
Whether we fit the trajectory with $\alpha_{2}$ only or together
with $\alpha_{3}$, $\alpha_{5}$, and $\alpha_{8}$, it is possible
to obtain a good estimation only for $\alpha_{2}$, the only parameter
that can be used for determine the depth. The other Slepian functions
can be used only to absorb other signals, preventing biases into $\alpha_{2}$.
Note that the uncertainty on $\alpha_{2}$ does not dramatically increase
when including also the other Slepian functions.

\begin{figure}
\centering{}\includegraphics[scale=0.35]{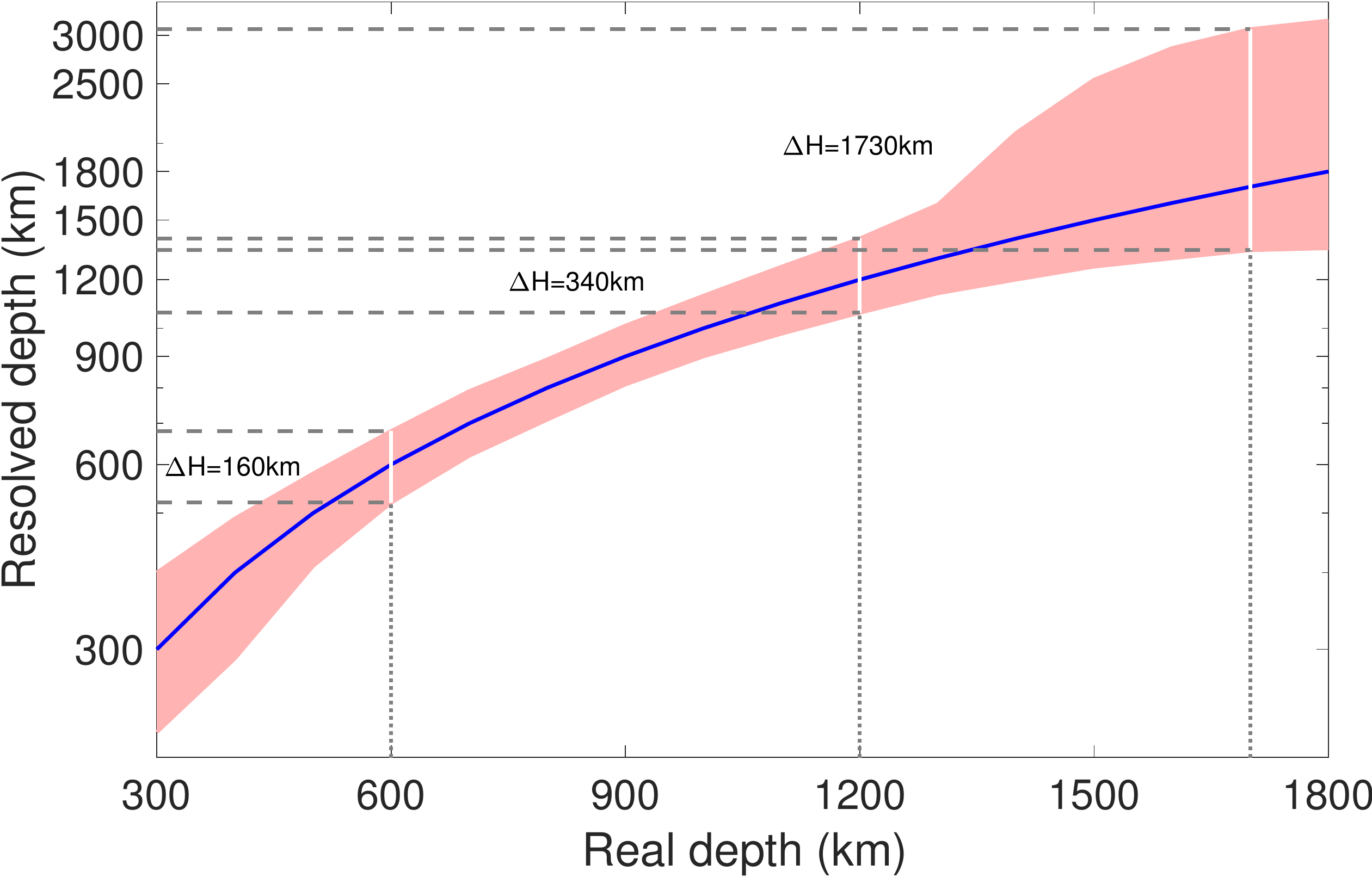}\caption{The detectability of the GRS depth. The uncertainty in the detection
of the GRS depth for a range of depths. The solid line denotes the
resolved depth as function of the real depth (a one to one connection).
The shaded region shows the uncertainty in the resolved depth calculated
from the uncertainty in the resolved $\alpha_{2}$. \label{fig: depth-detectibility}}
\end{figure}

We can now use these results to estimate the detectability of the
GRS with the upcoming Juno overflights and the Slepian approach. Given
that we are able to resolve $\alpha_{2}$ we examine how the uncertainty
associated with it is translated to an uncertainty in the estimated
depth. For each depth $H$, we take the TW value of $\alpha_{2}$
plus the uncertainty given with the TE solution $\delta\alpha_{2}$
and search for depth $H^{+}$ at which the TW value matches the combined
value, so that $\alpha_{2}(H^{+})=\alpha_{2}(H)+\delta\alpha_{2}(H)$.
The difference between the depths $\delta H^{+}=H^{+}-H$ is taken
as the upper uncertainty for the resolved depth. Similarly, we find
$\delta H^{-}$ using $\delta\alpha_{2}$ and $H^{-}$. Note that
$\delta H^{+}\ne-\delta H^{-}$ as $\alpha_{2}$ is not a linear function
of $H$. The resolved depths and the uncertainties associated with
them are shown in Figure~\ref{fig: depth-detectibility}. For GRS
depths of 300 to 1300 km the uncertainty is of the order of $\pm$100~km.
For GRS depths larger than 1300 km, the lower uncertainty is similar
but the upper uncertainty grows considerably, because the $\alpha_{2}$
value grows slower at these depths. For example, if the winds are
1700~km deep, the lower uncertainty will be around 200 km but the
upper uncertainty will be around 1500~km.

\section{Conclusion}

How deep is Jupiter's Great Red Spot? Although it has been observed
for a few centuries, little is known about its structure and dynamics
below its observed cloud-level. The Juno mission will soon provide
an opportunity to resolve this long standing question. The single
Juno flyby over the GRS (PJ7) to date was dedicated to the microwave
radiometer, which showed that it is at least a couple hundred kilometer
deep \citep{Li2017a}. The next flybys over the GRS, PJ18 and PJ21,
to be carried during 2019, will allow high-precision gravity measurements
that might be used to estimate how deep the GRS winds penetrate below
the cloud-level. This however is a challenging task since the GRS
is a small feature whose gravity signal is close to the detectability
levels.

Here we propose a new method to determine the depth of the GRS using
the upcoming gravity measurements, a dynamical flow model, and a Slepian
functions approach that enables an effective representation of the
wind-induced gravity signal, and an efficient determination of the
GRS depth given the limited expected measurements.

We show that the gravity signal induced by the GRS winds can be well
represented with a basis of Slepian functions, defined specifically
for the GRS region. It is found that one function ($\alpha_{2}$)
dominates the signal for shallow cases, and for deeper winds additional
2-3 functions are needed ($\alpha_{3}$, $\alpha_{5}$ and $\alpha_{8}$),
therefore only a few parameters are needed in order to resolve the
gravity signal induced by the GRS and hence the depth of its winds.

Using the Juno trajectory estimation model we examine our ability
to detect a range of wind depths. We find that for GRS wind depths
of 300 to 1300~km the methodology allows to resolve the depth of
the GRS winds with an accuracy of about $\pm$100~km. For GRS depths
larger than 1300 km, the lower uncertainty is similar but the upper
uncertainty grows considerably.

~

\textit{Acknowledgements:} This research has been supported by the
Israeli Ministry of Science, and the Helen Kimmel Center for Planetary
Science at the Weizmann Institute of Science. DD was  supported by
the Italian Space Agency (ASI).

\end{document}